\documentclass[twocolumn,pra,aps
reprint, superscriptaddress,
showpacs,
amsmath,amssymb,
aps, prl,
]{revtex4-1}
\usepackage{amssymb}
\usepackage{amsfonts}
\usepackage{amsmath}
\usepackage{cases}
\usepackage{stmaryrd}\usepackage{tipa}
\usepackage[dvips]{graphicx}
\usepackage{dcolumn}
\usepackage{bbold}
\usepackage{graphicx}
\usepackage{subfigure}
\usepackage{dcolumn}
\usepackage{bm}
\usepackage{color}
\usepackage{xcolor}
\usepackage{CJK}
\usepackage{subfigure}

\newcommand{\beq}{\begin{equation}}
\newcommand{\eeq}{\end{equation}}
\newcommand{\beqa}{\begin{eqnarray}}
\newcommand{\eeqa}{\end{eqnarray}}

\begin{document}

\title{Switchable particle statistics with an embedding quantum simulator}

\author{Xiao-Hang Cheng}
\affiliation{Department of Physics, Shanghai University, 200444 Shanghai, People's Republic of China}
\affiliation{Department of Physical Chemistry, University of the Basque Country UPV/EHU, Apartado 644, 48080 Bilbao, Spain}

\author{I\~{n}igo Arrazola}
\affiliation{Department of Physical Chemistry, University of the Basque Country UPV/EHU, Apartado 644, 48080 Bilbao, Spain}

\author{Julen S. Pedernales}
\affiliation{Department of Physical Chemistry, University of the Basque Country UPV/EHU, Apartado 644, 48080 Bilbao, Spain}

\author{Lucas Lamata}
\affiliation{Department of Physical Chemistry, University of the Basque Country UPV/EHU, Apartado 644, 48080 Bilbao, Spain}

\author{Xi Chen}
\affiliation{Department of Physics, Shanghai University, 200444 Shanghai, People's Republic of China}

\author{Enrique Solano}
\affiliation{Department of Physical Chemistry, University of the Basque Country UPV/EHU, Apartado 644, 48080 Bilbao, Spain}
\affiliation{IKERBASQUE, Basque Foundation for Science, Maria Diaz de Haro 3, 48013 Bilbao, Spain}
\date{\today}

\begin{abstract}
We propose the implementation of a switch of particle statistics with an embedding quantum simulator. By encoding both Bose-Einstein and Fermi-Dirac statistics into an enlarged Hilbert space,  the statistics of the simulated quantum particles may be changed {\it in situ} during the time evolution, from bosons to fermions and from fermions to bosons, as many times as desired before a measurement is performed. We illustrate our proposal with few-qubit examples, although the protocol is straightforwardly extendable to larger numbers of particles. This proposal can be implemented on different quantum platforms such as trapped ions, quantum photonics, and superconducting circuits, among others. The possibility to implement permutation symmetrization and antisymmetrization of quantum particles enhances the toolbox of quantum simulations, for unphysical operations as well as for symmetry transformations.
\end{abstract}


\maketitle

\begin{figure*}[]
{\includegraphics[width=1 \linewidth]{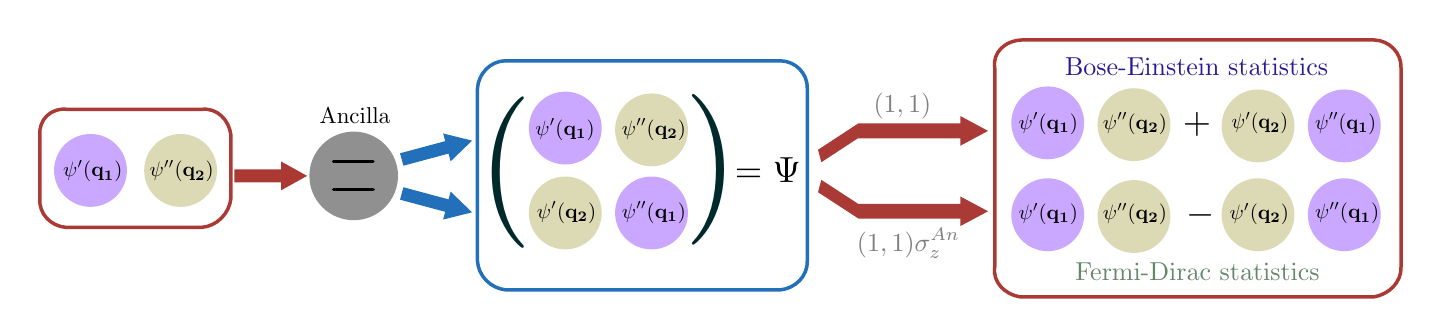}}
\vspace{-1 cm}
\caption{(color online) Scheme of the proposed embedding quantum simulator for switchable two-particle statistics.}
\end{figure*}
\vspace{-5 cm}

\section{Introduction}
In quantum mechanics, particles fall into two fundamental categories: bosons or fermions. Bosons, with integer spin and a wave function that is symmetric under the permutation of two particles, follow Bose-Einstein statistics, for example photons and gluons. Fermions, with half-integer spin, have a wave function that is antisymmetric under the exchange of two particles and are characterized by Fermi-Dirac statistics, as is the case of quarks and leptons. The investigation of permutations of quantum particles is a fundamental topic in many-body physics~\cite{many body}. Superselection rules establish that every quantum particle follows a well-determined quantum statistics during its whole lifespan. Is it possible to switch the particle statistics and to achieve the change of a quantum particle from a boson to a fermion or viceversa, during a quantum dynamics? We may answer this question in the positive, by means of a suitable encoding of the dynamics onto a quantum simulator.
\vspace{-0.1cm}

Since proposed by Richard Feynman \cite{Feynman}, the field of quantum simulation has strived to observe complex quantum phenomena in controllable quantum platforms. Some examples of the latter are trapped ions \cite{trapped ion, trapped ion2}, superconducting circuits \cite{superconducting}, ultracold quantum gases \cite{ultracold gas}, and photonic systems \cite{photonics}. Proposals and experiments of quantum simulations for relativistic quantum physics \cite{ qs dirac, nature, Klein tunneling and Dirac potentials, Klein Paradox, dirac particles, majorana, relativistic physics QED},  quantum chemistry \cite{quantum chemistry, quantum chemistry circuits}, spin systems \cite{spin system ions, Friedenauer08, ising spins, spin system circuits, spin system digital analog}, quantum field theories \cite{quantum field theories ions, quantum field theories fermion,ZollerQFT,Lloyd fermion,Wilhelm16}, and quantum phase transitions \cite{quantum phase transition}, among others, have already been achieved. Besides the standard unitary and dissipative evolutions, the paradigm of embedding quantum simulation allows for the possibility of realizing unphysical operations~\cite{majorana}. Encoding the simulated dynamics in an enlarged embedding space, antilinear and antiunitary evolutions \cite{majorana, SzameitMajo, KihwanMajo, embedding quantum simulator,embedding time correlation, embedding entanglements}, as well as noncausal kinematic transformations \cite{noncausal, time and spatial}, have been mapped onto physical processes.

In this work, we investigate the quantum simulation of a switch of particle statistics, from bosons to fermions and from fermions to bosons, reversibly, during a quantum evolution. We construct a spinor in an embedding space which has as components the wave functions of all different permutations of the particles, such that we can recover both the dynamics of Bose-Einstein and Fermi-Dirac statistics via a direct mapping. In this way, we can shift from bosonic to fermionic statistics and viceversa by means of local operations. The implementation of this proposal can be carried out via analog~\cite{qs dirac,nature} or digital \cite{digital QS, digital exp1, digital exp2, digital exp3} quantum simulations on a variety of quantum platforms.

\section{Theoretical Frame}
In the context of embedding quantum simulations, one has to distinguish between the simulated or embedded Hilbert space, which contains the dynamics of interest, and the simulating or embedding space, an enlarged Hilbert space containing the simulated model under a suitable mapping. This embedding space will allow for operations that under the considered mapping correspond to a process of interest in the simulated space, in this case, a modification in the particle statistics of the simulated model. 

For clarity, we will first illustrate our proposal considering the case of two particles, and later show that the extension to $N$ particles is polynomial in terms of resources. In order to encode bosonic and fermionic statistics in the embedding space, we define an enlarged spinor in the following way,
\vspace{-0.2 cm}
\begin{equation}
\label{spinor}
\Psi=\frac{1}{\sqrt{2}}\left(\begin{array}{cc}\psi'(q_1)\psi''(q_2) \\ \psi'(q_2)\psi''(q_1)\end{array}\right),
\vspace{-0.1 cm}
\end{equation}
where $\{\psi^{(j)}(q_i)\}_{j=1\cdots N}$ denotes a set of orthonormal wave functions for particle $i$, with $q_i$ representing all its degrees of freedom. We consider an orthonormal set in the fermionic case without loss of generality, given that linearly dependent contributions will cancel out due to antisymmetrization. In the bosonic case, non-orthogonal contributions could be equivalently encoded. The enlarged spinor, that we name {\it symmetrization spinor}, may be constructed via additional quantum levels or an auxiliary qubit. By this encoding, Bose-Einstein and Fermi-Dirac statistics can be recovered by an appropriate mapping,
\vspace{-0.15 cm}
\begin{eqnarray}
\psi_b&=&(1, 1)\Psi=\frac{1}{\sqrt{2}}[\psi'(q_1)\psi''(q_2)+\psi''(q_1)\psi'(q_2)], \label{EQS1}  \\
\psi_f&=&(1, 1)\sigma_z\Psi=\frac{1}{\sqrt{2}}[\psi'(q_1)\psi''(q_2)-\psi''(q_1)\psi'(q_2)].\label{EQS2} \nonumber
\end{eqnarray}
To physically implement such a mapping, we encode the spinor by means of an ancillary qubit such that\vspace{0.05cm} ${\Psi= \frac{1}{\sqrt{2}}[|\!\uparrow\rangle\otimes \psi'(q_1) \psi''(q_2) + |\! \downarrow \rangle\otimes \psi'(q_2) \psi''(q_1)]}$, where $|\!\uparrow\rangle\equiv{1\choose0}$ and $|\!\downarrow\rangle \equiv{0\choose1}$. Then, with a local operation on the ancilla we can generate a symmetric wavefunction associated with the ancillary state $|\!\uparrow \rangle$ and an antisymmetric one associated with the state of the ancilla $|\! \downarrow \rangle$,
\begin{equation}
\label{spinor2}
\frac{1}{\sqrt{2}}(\sigma_x+\sigma_z)\Psi=\frac{1}{\sqrt{2}}{\psi_b\choose \psi_f}.
\end{equation}
Measuring the ancilla and postselecting either states $|\!\uparrow \rangle$ or $|\! \downarrow \rangle$, we will be left with the system in a bosonic or fermionic state, respectively.  We point out that in order to switch from bosonic to fermionic statistics of the simulated particles during a quantum dynamics, all that is needed is to apply a local $\sigma_z$ operation on the enlarged spinor in the embedding space, before implementing the mapping to the bosonic or fermionic state.

We assume now that the evolution in the simulated space is ruled by a Hamiltonian $H$, which typically will be permutation invariant to preserve the particle statistics. Thus, the corresponding Hamiltonian in the embedding space, which preserves the mapping between the enlarged spinor and the fermionic and bosonic wave functions at all times, is given by ${\tilde H= \mathbb 1_2 \otimes H}$. This makes our proposal specially suitable for a quantum simulation, as the dynamics of any demonstrated quantum simulator will remain the same in the enlarged space. All that is needed is to add an ancilla qubit in the appropriate entangled initial state, and then  keep it out of the evolution. In trapped ions, this can be done via individual addressing, using ions of different species that will not interact with the lasers that rule the evolution of the system, or by changing the energy splitting of the ionic qubits to bring them out of resonance~\cite{KihwanMajo}. In circuit QED, superconducting qubits can also be taken out of resonance with the cavity to isolate them from the dynamics of the system. This is typically done by imposing a magnetic field across the superconducting qubit that will change the energy splitting of its two levels~\cite{superconducting}. If the Hamiltonian is not permutational invariant, ${H_{12} \neq H_{21}}$, then the enlarged Hamiltonian would be ${\tilde H= |\! \uparrow \rangle \langle\! \uparrow| \otimes H_{12} + |\! \downarrow \rangle \langle\! \downarrow | \otimes H_{21}}$, where ${|\! \uparrow \rangle}$ and ${|\! \downarrow \rangle}$ are the ancilla states corresponding to permutations $\{1,2\}$ and $\{2,1\}$ of the particles, respectively, and $H_{ij}$ the Hamiltonian corresponding to permutation $ij$.

In order to measure a given observable $M$ in the simulated quantum system, one can physically implement the mapping from the embedding space to the embedded one, as explained above, and then perform the measurement of interest. However, a more direct approach would be to encode the expectation value of the observable in the simulated space into the  expectation value of an observable in the enlarged Hilbert space. Depending on whether we are interested in the observable to be evaluated, according to bosonic or fermionic statistics, we will associate our observable to a different one in the enlarged space. Following the mappings in Eq.~(\ref{EQS1}), we have that
\vspace{-0.2 cm}
\begin{eqnarray}
\langle M \rangle_{\psi_b}&=&\langle \psi_b | M | \psi_b\rangle=\langle \Psi | {1\choose1} M (1, 1) |\Psi\rangle \nonumber \\
&=&\langle \Psi | (\mathbb{1}_2 + \sigma_x) \otimes M | \Psi\rangle,\nonumber\\
\langle M \rangle_{\psi_f}&=&\langle \psi_f | M | \psi_f\rangle=\langle \Psi | \sigma_z {1\choose1} M (1, 1) \sigma_z |\Psi\rangle \nonumber \\
&=&\langle \Psi | (\mathbb{1}_2 - \sigma_x) \otimes M | \Psi\rangle,\nonumber\\
\langle M \rangle_{\psi_b, \psi_f}&=&\langle \psi_b | M | \psi_f\rangle=\langle \Psi | {1\choose1} M (1, 1) \sigma_z |\Psi\rangle \nonumber \\
&=&\langle \Psi | (\sigma_z - i\sigma_y) \otimes M | \Psi\rangle.
\end{eqnarray}
In this way, one can measure not only the observable $M$ with the corresponding bosonic $\psi_b$ or fermionic $\psi_f$ wave function, but also correlations between these two kinds of particles. These correlations will cancel out for standard permutation-invariant measurement operators, due to superselection rules. However, they can be a useful consistency check in a quantum simulation implementation, or they can give nonzero results when observables that connect both types of statistics are used, e.g., non-permutation-invariant ones. Our method has as a by-product the capability to test permutation invariance of an unknown Hamiltonian by detecting correlations between bosonic and fermionic sectors.

\section{Extension to $N$ Particles}

\begin{figure*}[]
{\includegraphics[width=1 \linewidth]{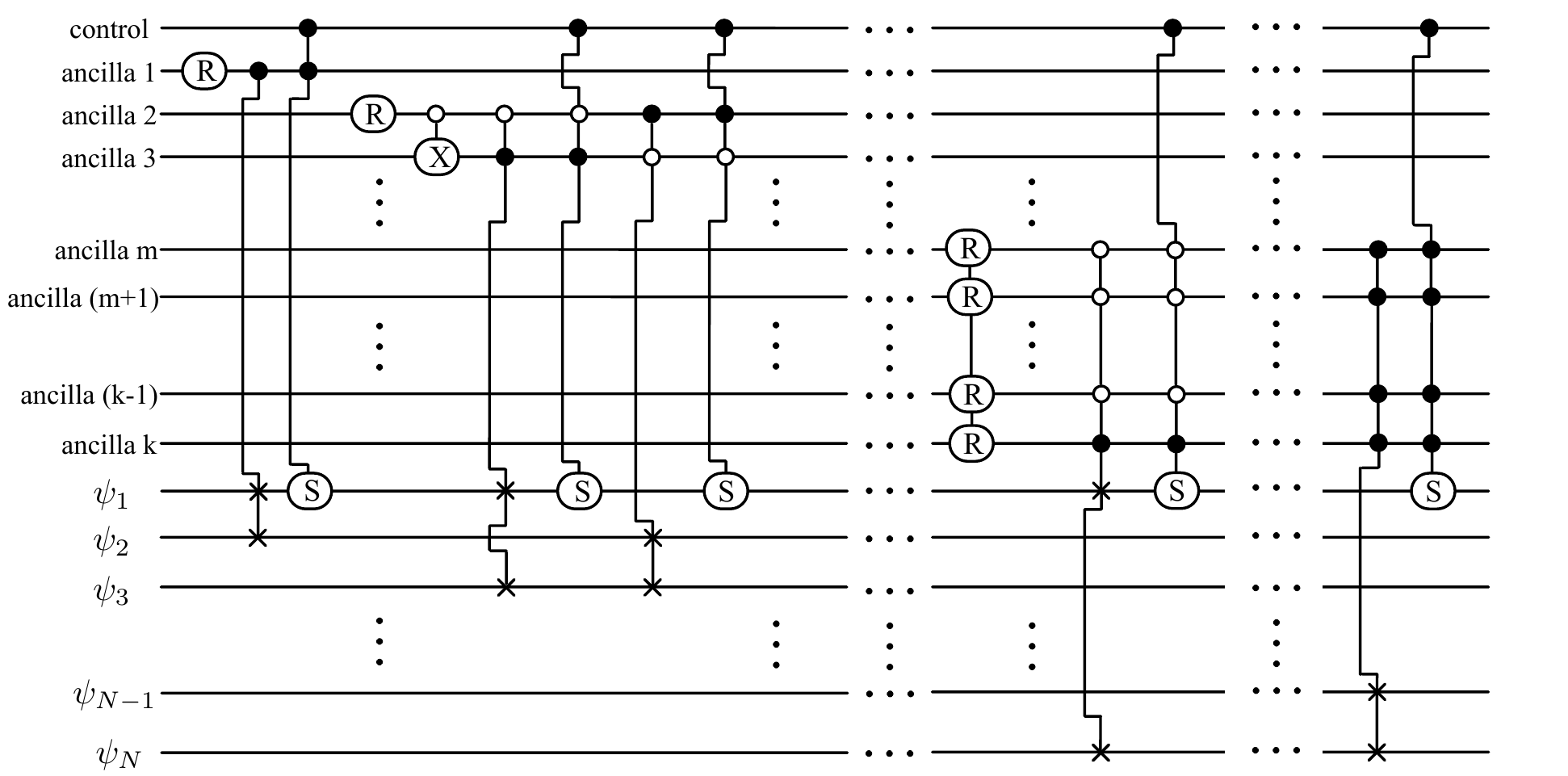}}
\caption{Scheme of the proposed protocol for N-particle statistics when $\log_2 N$ is a natural number. Rotation gate $R=\exp(-i \sigma_y \pi/4)$. S gate is used to introduce the minus sign. And white and black control dot stands for spin up and down, respectively.}
\end{figure*}

For an $N$-particle system one can proceed similarly, composing the enlarged spinor in the embedding Hilbert space, i.e., the symmetrization spinor, with the permutation states of the particles. To generate such a spinor we consider an algorithm that makes use of a generic controlled-swap gate that will permute the wave functions of any two particles. This kind of gates has been recently performed in the laboratory~\cite{expSWAPgate}. The algorithm consists of $N-1$ steps, $N$ being the total number of particles being simulated. At the beginning the system is in state $\psi'(q_1)\psi''(q_2)\cdots \psi^{(N)}(q_N)$. We introduce an ancillary qubit in a superposition of $|\!\uparrow\rangle$ and $|\!\downarrow\rangle$ states, and apply a controlled-swap gate, which will swap the particles $1$ and $2$ only if the ancillary qubit is in state $|\!\downarrow\rangle$, that is $U_{\rm S}= |\!\uparrow\rangle\langle\!\uparrow| \otimes \mathbb{1} + |\!\downarrow\rangle\langle\!\downarrow|\otimes S_{12}$, where $S_{ij}\psi'(q_1)\psi''(q_2)\cdots \psi^{(i)}(q_i) \cdots \psi^{(j)}(q_j) \cdots \psi^{(N)}(q_N)= \psi'(q_1)\psi''(q_2)\cdots \psi^{(i)}(q_j) \cdots \psi^{(j)}(q_i) \cdots \psi^{(N)}(q_N)$. This will generate the spinor
\begin{eqnarray}
\Psi= \frac{1}{\sqrt{2}}\left(\begin{array}{ccccc}\psi'(q_1)\psi''(q_2)\cdots \psi^{(N)}(q_N) \\ \psi'(q_2) \psi''(q_1) \cdots \psi^{(N)}(q_N) \end{array}\right).
\end{eqnarray}
The second step of the algorithm will need to introduce two more ancillary qubits to accommodate all the three new permutations allowed by particle 3. We shall initialize the two qubits in a vector state with only three nonzero componentes, that is, we will set the coefficient of state $|\!\downarrow\!\downarrow\rangle$ to zero, while the rest of states will be in an even superposition. Associated with state $|\!\uparrow\!\uparrow\rangle$, we will have the spinor containing the permutations of particles $1$ and $2$. We continue by permuting particles 1 and 3 controlled on state $ |\!\uparrow\!\downarrow\rangle$, and particles $2$ and $3$ controlled on the state $ |\!\downarrow\!\uparrow\rangle$. This will generate a spinor containing all the possible permutations between particles 1 to 3, and also some zeros. We will proceed in a similar fashion employing in each step a number of qubits enough to accommodate the $n$ new permutations that each $n$th particle introduces.  These qubits will be first initialized in an even superposition of $n$ and only $n$ orthogonal states, which can always be done efficiently~\cite{Long01}.  The state $|\!\uparrow\!\uparrow\cdots\!\uparrow\rangle$ of the set of qubits introduced in the $n$th step will always be left associated with the already constructed spinor, containing the permutations of the $n-1$ previous particles. We will then perform a permutation between every previous particle and the $n$th one, via $n$ swap gates each controlled over one of the available states of the qubits introduced in this step of the protocol. For the antisymmetrized state one will proceed similarly but introducing a minus sign additional to each swap gate. By an auxiliary control qubit $q_c$, both processes for the bosonic and fermionic statistics can be performed in parallel, via a controlled sign added or not to each controlled-swap gate. This will be done depending on the simulated statistics, either fermionic or bosonic. After $N-1$ steps, we will have constructed a spinor containing all possible permutations both for bosonic and fermionic cases, and some zeros.

Notice that while the total number of permutations per case, bosonic or fermionic, is $N!$, we have only made use of $n-1$ controlled-swap gates per step $n$, that is a total number of gates of $(N-1)(N-2)/2$. This is because in each step $n$, the swap gate acts over the whole spinor containing the permutations of all previous $n$-1 particles. In this way, we profit from the accumulated permutations, making the algorithm polynomial in terms of the number of gates. The total number of added ancillary qubits grows as $\kappa\equiv \log_2 2+ \overline{\log_2 3} + \log_2 4 + ...< N^2$, where the upper bar indicates that we round up to the first natural number. As a consequence, the number of ancillary qubits grows subquadratically with $N$. One must take into account that the controlled-swap gates need to be controlled over a higher number of qubits in each step. However, it is known that any unitary operation controlled over a single qubit can be efficiently extended to a unitary controlled over $n$ qubits by the addition of $n-1$ working qubits and $2(n-1)$ Toffolli gates~\cite{Nielsen&Chuang}. On the other hand, swap gates of arbitrary dimension or even continuous  degrees of freedom have already been considered in the literature~\cite{Garcia-Escartin13, Wang01}.

After finishing the initialization protocol, we will have the $N!$-component symmetrization spinors containing all permutations both for bosonic (symmetric) and fermionic (antisymmetric) cases, entangled with the qubit $q_c$ that controls the statistics,
\begin{eqnarray}
|\Psi\rangle= \frac{1}{\sqrt{2N!}}\left[|\!\uparrow\rangle_c|\Psi_{\rm sym} \rangle+|\!\downarrow\rangle_c|\Psi_{\rm asym} \rangle\right]. 
\end{eqnarray}
One can then evolve this state with the embedding Hamiltonian, $\tilde{H}=\mathbb 1 \otimes H$, where $\mathbb 1$ acts on the Hilbert space of the enlarged symmetrization spinor and of $q_c$, and $H$ acts on the simulated system. This evolution can be implemented straightforwardly and will respect the particle statistics when considering permutationally invariant $H$ dynamics. At the end of the quantum simulation, we will obtain the relevant information, either for bosonic or fermionic state, according to the following procedure.  We will measure the statistics control qubit, $q_c$, and postselect for a bosonic result if the outcome is $s=\uparrow$, or fermionic result if the outcome is $s=\downarrow$, with 1/2 probability. Then, we propose to measure the observable $\tilde{M}=(\sigma_x+\mathbb 1_2)^{\otimes \kappa}\otimes M$, where $(\sigma_x+\mathbb 1_2)^{\otimes \kappa}$ acts on the degree of freedom of the symmetrization spinor, and can be mapped onto $\kappa$ local $\sigma_z$ measurements via a local unitary on each of the $\kappa$ auxiliary qubits and a constant shift of each local outcome by 1. This observable $\tilde{M}$ in the enlarged space corresponds to observable $M$ in the simulated space, namely, 
\begin{equation}
\langle \Psi | s=\uparrow, \downarrow\rangle_c\langle s=\uparrow, \downarrow\! | \tilde{M} |\Psi\rangle=\langle \psi_{b,f}|M|\psi_{b,f}\rangle/2,
\end{equation}
where the bosonic(fermionic) wave function $|\psi_b\rangle$($|\psi_f\rangle$) is given in terms of the corresponding symmetrization spinor as $|\psi_{b,f}\rangle=(1,1,...,1)|\Psi_{\rm sym,asym}\rangle$.
Therefore, with a single implementation, one can obtain the measurement outcome for both the fermionic and bosonic cases, postselected on the statistics control qubit. One may also measure cross correlation between fermionic and bosonic statistics, which will cancel out for permutationally invariant $H$ due to the superselection rule, but can serve as a consistency check in the quantum simulation. This may be done via the measurement of the observable $\sigma_x\otimes \tilde{M}$ in the enlarged space, where $\sigma_x$ acts on $q_c$, which in this case is not postselected. We point out that, while the correspondence between the expectation values of the operators $\tilde M$ in the embedding space and $M$ in the simulated space is direct, the same does not hold true for higher moments of the operator $\tilde M$. For instance 
\begin{equation}
\langle \Psi | s=\uparrow, \downarrow\rangle_c\langle s=\uparrow, \downarrow\! | \tilde{M}^2 |\Psi\rangle=2^{k -1}\langle \psi_{b,f}|M^2|\psi_{b,f}\rangle.
\end{equation}
The direct consequence of this is that the variance of our observable in the embedding space grows exponentially with the number of simulated particles, forcing us to enlarge the size of the sample of measurements in order to preserve the accuracy of the retrieved expectation value. This fact will as well set a limitation on the accuracy to which correlations between fermions and bosons can be measured, and therefore the precision to which the permutation invariance of an unknown Hamiltonian can be tested. However, for a small enough number of particles our method is still feasible, and could shed light on physically relevant problems.

\section{Particle Statistics Switch Cases}
We include for illustrative purposes an extensive analysis of two- and three-particle cases, as well as of the switching between Bose-Hubbard and Fermi-Hubbard models via our techniques. These cases can be implemented via purely analog or digital-analog techniques~\cite{MS gate NJP, MS gate}.
\subsection{Spin Dynamics with Exchange Hamiltonian}
As a first example, we consider purely spin dynamics, consisting of the exchange coupling Hamiltonian acting on a two-particle system,
\begin{equation}
\label{Exchange Hamiltonian}
H=g(\sigma_1^+ \otimes \sigma_2^- + \sigma_1^- \otimes \sigma_2^+).
\end{equation}
Accordingly, the simulating Hamiltonian acting on the enlarged symmetrization spinor in the embedding Hilbert space reads
\begin{equation}
\tilde{H}=\mathbb{1}_2 \otimes H=g(\mathbb{1}_2 \otimes \sigma_1^+ \otimes \sigma_2^- + \mathbb{1}_2 \otimes \sigma_1^- \otimes \sigma_2^+).
\end{equation} Physically this would correspond to having two qubits evolving under the exchange Hamiltonian in Eq.~(\ref{Exchange Hamiltonian}) while a third qubit is left untouched. This kind of Hamiltonians has already been implemented in ions~\cite{exchange Hamiltonian} and in superconducting circuits~\cite{exchange Hamiltonian circuits}, among others.
For the two-particle initial state 
\begin{eqnarray}
\Psi_0=\frac{1}{\sqrt{2}}\left(\begin{array}{cc}|\!\uparrow\rangle_1 |\!\downarrow\rangle_2 \\ |\!\downarrow\rangle_1 |\!\uparrow\rangle_2 \end{array}\right),
\end{eqnarray}
the time evolution of the embedding spinor is given by
\begin{eqnarray}
\Psi(t)&=&e^{-i \tilde{H} t}\Psi_0=e^{-i g t (\mathbb{1}_2 \otimes \sigma_1^+ \otimes \sigma_2^- + \mathbb{1}_2 \otimes \sigma_1^- \otimes \sigma_2^+)}\Psi_0\nonumber\\
&=&\frac{1}{\sqrt{2}}\left(\begin{array}{cc}\cos(g t)|\!\uparrow\rangle_1|\!\downarrow\rangle_2-i \sin(g t)|\!\downarrow\rangle_1|\!\uparrow\rangle_2 \\ -i \sin(g t)|\!\uparrow\rangle_1|\!\downarrow\rangle_2+\cos(g t)|\!\downarrow\rangle_1|\!\uparrow\rangle_2 \end{array}\right).\nonumber\\
\end{eqnarray}
Thus, one can obtain the bosonic and the fermionic wave functions applying the corresponding mappings, and has the ability to switch between both simulated statistics via a $\sigma_z$ gate acting on the enlarged spinor,
\begin{eqnarray}
\psi_b&=&(1,1)\Psi(t)=\frac{e^{-i g t}}{\sqrt{2}}(|\!\uparrow\rangle_1|\!\downarrow\rangle_2+|\!\downarrow\rangle_1|\!\uparrow\rangle_2), \nonumber \\
\psi_f&=&(1,1)\sigma^z\Psi(t)=\frac{e^{i g t}}{\sqrt{2}}(|\!\uparrow\rangle_1|\!\downarrow\rangle_2-|\!\downarrow\rangle_1|\!\uparrow\rangle_2).
\end{eqnarray}

We consider the observable ${M=\sigma_1^x \otimes \sigma_2^x}$ in the simulated space, which has as corresponding observables in the enlarged space  $\tilde M_b= (\mathbb 1_2 + \sigma^x)\otimes \sigma_1^x \otimes \sigma_2^x$, for bosons, and $\tilde M_f= (\mathbb 1_2 - \sigma^x)\otimes \sigma_1^x \otimes \sigma_2^x$, for fermions, as well as ${\tilde M_{bf}= (\sigma^z - i\sigma^y)\otimes \sigma_1^x \otimes \sigma_2^x}$, for their correlation. The expectation values of these observables are time independent and have different values for all the three cases considered here,
\begin{eqnarray}
&&\langle \tilde M_b \rangle_{\Psi}=\langle M \rangle_{\psi_b}=1, \nonumber\\
&&\langle \tilde M_f \rangle_{\Psi}=\langle M \rangle_{\psi_f}=-1, \nonumber\\
&&\langle \tilde M_{bf} \rangle_{\Psi}=\langle M \rangle_{\psi_b, \psi_f}=0.
\end{eqnarray}
We can therefore differentiate via $M$ the bosonic and fermionic wave function dynamics, and their correlation vanishes, as expected, due to the superselection rule.

\subsection{Spin Dynamics with Heisenberg Hamiltonian}
Here we consider the dynamics associated with the Heisenberg spin Hamiltonian in our simulated space,
\begin{equation}
H^H=g(\sigma_1^x \otimes \sigma_2^x + \sigma_1^y \otimes \sigma_2^y + \sigma_1^z \otimes \sigma_2^z).
\end{equation}
The corresponding simulating Hamiltonian in the embedding space is
\begin{equation}
\tilde{H}^H=\mathbb{1}_2 \otimes H^H=g(\mathbb{1}_2 \otimes \sigma_1^x \otimes \sigma_2^x+\mathbb{1}_2 \otimes \sigma_1^y \otimes \sigma_2^y+\mathbb{1}_2 \otimes \sigma_1^z \otimes \sigma_2^z).
\end{equation}
We assume that the initial state for two particles in the enlarged Hilbert space is
\begin{eqnarray}
\Psi_0=\frac{1}{\sqrt{2}}\left(\begin{array}{cc}|\!\uparrow\rangle_1 |\!\downarrow\rangle_2 \\ |\!\downarrow\rangle_1 |\!\uparrow\rangle_2 \end{array}\right).
\end{eqnarray}
After the time evolution, the wave function in the enlarged Hilbert space is
\begin{equation}
\Psi(t)=\frac{1}{\sqrt{2}}\left(\begin{array}{cc}e^{i g t}\cos(2 g t)|\!\uparrow\rangle_1 |\!\downarrow\rangle_2-i e^{i g t}\sin(2 g t)|\!\downarrow\rangle_1 |\!\uparrow\rangle_2 \\ -i e^{i g t}\sin(2 g t)|\!\uparrow\rangle_1 |\!\downarrow\rangle_2+e^{i g t}\cos(2 g t)|\!\downarrow\rangle_1 |\!\uparrow\rangle_2\end{array}\right).
\end{equation}
Therefore, the expressions of Bose-Einstein and Fermi-Dirac statistics are,
\begin{eqnarray}
\psi_b&=&(1,1)\Psi(t)=\frac{e^{-i g t}}{\sqrt{2}}(|\!\uparrow\rangle_1|\!\downarrow\rangle_2+|\!\downarrow\rangle_1|\!\uparrow\rangle_2), \\
\psi_f&=&(1,1)\sigma_z\Psi(t)=\frac{e^{3i g t}}{\sqrt{2}}(|\!\uparrow\rangle_1|\!\downarrow\rangle_2-|\!\downarrow\rangle_1|\!\uparrow\rangle_2).
\end{eqnarray} 
Similarly, we can obtain the expectation values of observable $M=\sigma_1^x \otimes \sigma_2^x$ for both Bose-Einstein and Fermi-Dirac dynamics, as well as for their correlations,
\begin{eqnarray}
\langle M\rangle_{\psi_b}=1, \langle M \rangle_{\psi_f}=-1, \langle M \rangle_{\psi_b, \psi_f}=0.
\end{eqnarray} 

\subsection{Motional-Spin Dynamics (Jaynes-Cummings)}
We consider now as an illustrative example a system of two particles containing spins and continuous degrees of freedom. The dynamics will be given by the resonant Jaynes-Cummings Hamiltonian in the simulated space, $H_{JC}=g\sum\nolimits_{i=1}^{2}\left(a_i\sigma_i^{+}+a_i^{\dag}\sigma_i^{-}\right)$, where $g$ is the coupling strength, $a_i$ and $a_i^{\dag}$ are the annihilation and creation operators acting on the continuous degrees of freedom of the system, and ${\sigma_i^{+,-}=(\sigma_i^x\pm i\sigma_i^y)/2}$ are the spin raising and lowering operators. The corresponding Hamiltonian in the enlarged space is ${\tilde{H}_{\rm{JC}}=\mathbb{1}_2 \otimes H_{\rm{JC}}}$. Jaynes-Cummings Hamiltonians are natural in superconducting circuits~\cite{JC circuits}, and can be generated by means of red-sideband interactions in trapped ions~\cite{trapped ion}.

We take as initial state
\begin{eqnarray}
\psi_1(x_1,t=0)=|n\rangle_1|\!\uparrow\rangle_1, \psi_2(x_2,t=0)=|m\rangle_2|\!\downarrow\rangle_2, \nonumber\\
\end{eqnarray}
where $\psi_1$ and $\psi_2$ represent the initial states of particles $1$ and $2$ before symmetrization or antisymmetrization, with $|\!\uparrow\rangle$ and $|\!\downarrow\rangle$ denoting the spin state, and $|n\rangle$ being a Fock state associated with the continuous degree of freedom of each particle. The corresponding state for the enlarged spinor reads
\begin{eqnarray}
\Psi_0=\frac{1}{\sqrt{2}}\left(\begin{array}{cc}|n\rangle_1|\!\uparrow\rangle_1\otimes|m\rangle_2|\!\downarrow\rangle_2 \\ |m\rangle_1|\!\downarrow\rangle_1\otimes|n\rangle_2|\!\uparrow\rangle_2\end{array}\right).
\end{eqnarray}
Therefore, the bosonic and fermionic wave functions after the time evolution are given by,
\begin{widetext}
\begin{eqnarray}
\psi_b(t)=&&\frac{1}{\sqrt{2}}\left[\cos(gt\sqrt{n+1})\cos(gt\sqrt{m})|n\rangle_1|m\rangle_2-\sin(gt\sqrt{m})\sin(gt\sqrt{n+1})|m-1\rangle_1|n+1\rangle_2\right]|\!\uparrow\rangle_1|\!\downarrow\rangle_2\nonumber\\
&&-\frac{1}{\sqrt{2}}\left[\sin(gt\sqrt{n+1})\sin(gt\sqrt{m})|n+1\rangle_1|m-1\rangle_2-\cos(gt\sqrt{m})\cos(gt\sqrt{n+1})|m\rangle_1|n\rangle_2\right]|\!\downarrow\rangle_1|\!\uparrow\rangle_2\nonumber\\
&&-\frac{i}{\sqrt{2}}\left[\sin(gt\sqrt{n+1})\cos(gt\sqrt{m})|n+1\rangle_1|m\rangle_2+\cos(gt\sqrt{m})\sin(gt\sqrt{n+1})|m\rangle_1|n+1\rangle_2\right]|\!\downarrow\rangle_1|\!\downarrow\rangle_2\nonumber\\
&&-\frac{i}{\sqrt{2}}\left[\cos(gt\sqrt{n+1})\sin(gt\sqrt{m})|n\rangle_1|m-1\rangle_2+\sin(gt\sqrt{m})\cos(gt\sqrt{n+1})|m-1\rangle_1|n\rangle_2\right]|\!\uparrow\rangle_1|\!\uparrow\rangle_2, \nonumber\\
\psi_f(t)=&&\frac{1}{\sqrt{2}}\left[\cos(gt\sqrt{n+1})\cos(gt\sqrt{m})|n\rangle_1|m\rangle_2+\sin(gt\sqrt{m})\sin(gt\sqrt{n+1})|m-1\rangle_1|n+1\rangle_2\right]|\!\uparrow\rangle_1|\!\downarrow\rangle_2\nonumber\\
&&-\frac{1}{\sqrt{2}}\left[\sin(gt\sqrt{n+1})\sin(gt\sqrt{m})|n+1\rangle_1|m-1\rangle_2+\cos(gt\sqrt{m})\cos(gt\sqrt{n+1})|m\rangle_1|n\rangle_2\right]|\!\downarrow\rangle_1|\!\uparrow\rangle_2\nonumber\\
&&-\frac{i}{\sqrt{2}}\left[\sin(gt\sqrt{n+1})\cos(gt\sqrt{m})|n+1\rangle_1|m\rangle_2-\cos(gt\sqrt{m})\sin(gt\sqrt{n+1})|m\rangle_1|n+1\rangle_2\right]|\!\downarrow\rangle_1|\!\downarrow\rangle_2\nonumber\\
&&-\frac{i}{\sqrt{2}}\left[\cos(gt\sqrt{n+1})\sin(gt\sqrt{m})|n\rangle_1|m-1\rangle_2-\sin(gt\sqrt{m})\cos(gt\sqrt{n+1})|m-1\rangle_1|n\rangle_2\right ]|\!\uparrow\rangle_1|\!\uparrow\rangle_2.
\end{eqnarray} 
\end{widetext}
For the specific case of  Gaussian wave functions as initial states of the continuous degrees of freedom, we have ${\psi_1(x_1,s_1,t=0)=|\!\uparrow0\rangle_1}$, ${\psi_2(x_2,s_2,t=0)=|\!\downarrow0\rangle_2}$. Here, we have made use of the fact that Fock state $| 0 \rangle$ is a Gaussian distribution centered in the origin of the phase space. State ${|\!\downarrow0\rangle}$ is the ground state of the Jaynes-Cummings Hamiltonian, while state ${|\!\uparrow0\rangle}$ will evolve to generate the well-known Rabi oscillations,  ${\cos (gt) |\!\uparrow0\rangle-i\sin (gt) |\!\downarrow1\rangle}$. Therefore, the time evolution of the enlarged spinor in the embedding space is given by
\begin{equation}
\Psi(t)=\frac{1}{\sqrt{2}}\left(\begin{array}{cc}\cos (gt) |\!\uparrow 0 \rangle_1|\!\downarrow 0 \rangle_2-i\sin (gt) |\!\downarrow 1 \rangle_1|\!\downarrow 0 \rangle_2 \\ \cos (gt) |\!\downarrow 0 \rangle_1|\!\uparrow 0 \rangle_2-i\sin (gt) |\!\downarrow 0 \rangle_1|\!\downarrow 1 \rangle_2 \end{array}\right).
\end{equation}

Finally, one obtains bosonic and fermionic wave functions under the corresponding mapping, and with the ability to switch between both of the simulated statistics via a $\sigma_z$ gate acting on the enlarged spinor,
\begin{eqnarray}
\psi_b&=&(1,1)\Psi(t)\nonumber\\
&=&\frac{1}{\sqrt{2}}[\cos (gt) |\!\uparrow 0 \rangle_1|\!\downarrow 0 \rangle_2 - i\sin (gt) |\!\downarrow 1 \rangle_1|\!\downarrow 0 \rangle_2 \nonumber\\
 &&+ \cos (gt) |\!\downarrow 0 \rangle_1| \!\uparrow 0 \rangle_2 - i \sin (gt) |\!\downarrow 0 \rangle_1 |\!\downarrow 1 \rangle_2], \nonumber \\
\psi_f&=&(1,1)\sigma_z\Psi(t)\nonumber\\
&=&\frac{1}{\sqrt{2}}[\cos (gt) |\!\uparrow 0 \rangle_1 |\!\downarrow 0 \rangle_2  - i \sin (gt) |\!\downarrow 1 \rangle_1 |\!\downarrow  0 \rangle_2 \nonumber\\
&&- \cos (gt) |\!\downarrow 0 \rangle_1 |\!\uparrow 0 \rangle_2 + i \sin (gt) |\!\downarrow 0 \rangle_1 |\!\downarrow 1  \rangle_2].
\end{eqnarray}

\begin{figure}[]
\centering
\includegraphics[width=1.05\linewidth]{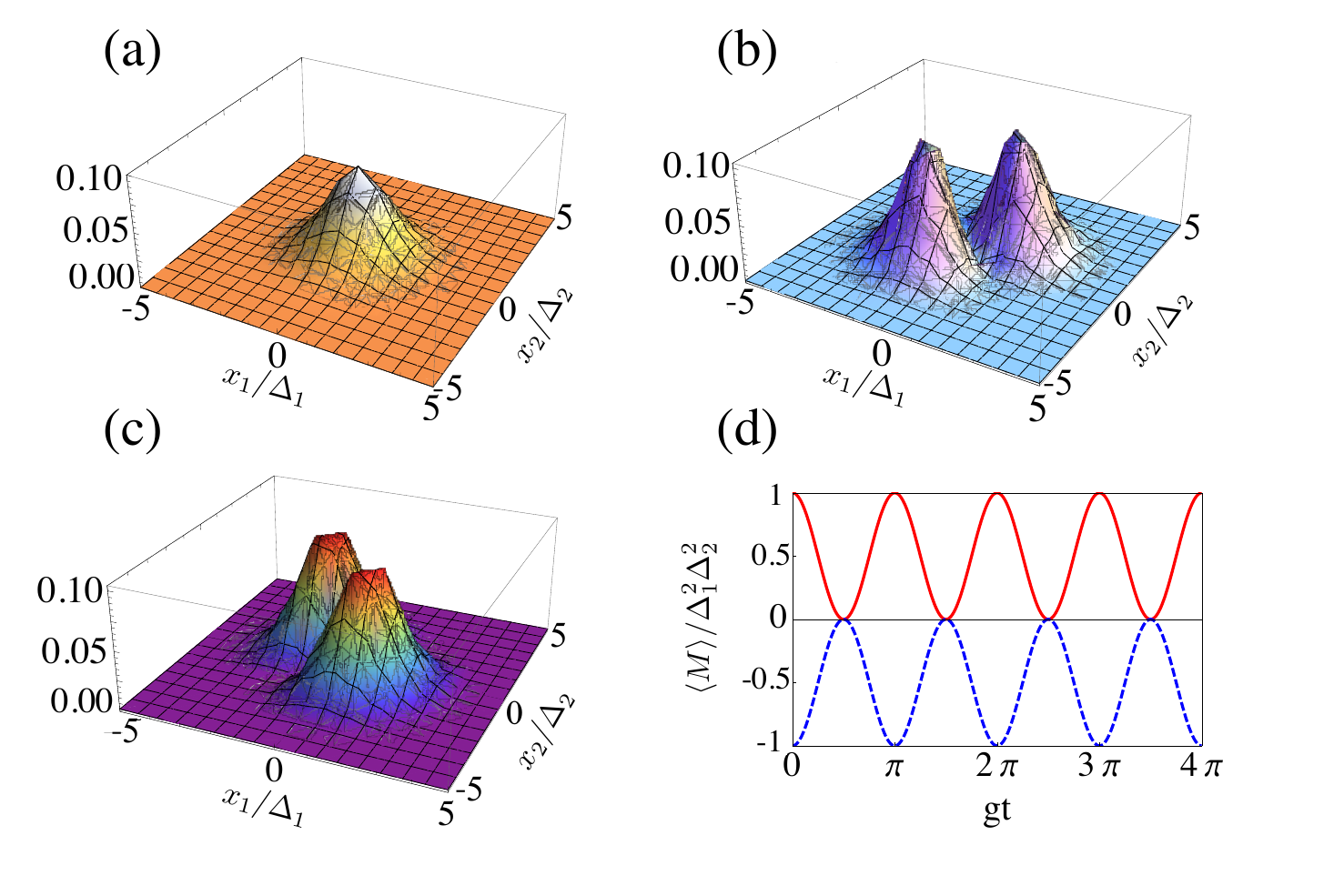}
\caption{(color online) Probability distributions $|\psi(x_1,x_2)|^2$ at time (a) $t=0$ for initial Gaussian wavepackets, and $t=\pi/2g$ for (b) Bose-Einstein and (c) Fermi-Dirac statistics, where $x_1$ and $x_2$ are in units of variances of the distributions $\Delta_1$ and $\Delta_2$, respectively. (d) Time-dependent expectation value for Bose-Einstein (solid red line) and Fermi-Dirac (dashed blue line) statistics of the nontrivial correlation ${M=x_1^2x_2^2\sigma_1^x\sigma_2^x}$.}
\label{JC}
\end{figure}

We plot in Fig.~\ref{JC}b-\ref{JC}c the bosonic/fermionic wave functions associated with the continuous degree of freedom of the system for $\cos(gt)=0$, i.e., $gt=\pi/2+s\pi$ with integer $s$. One can appreciate that the fermionic state cancels at the diagonal $x_1=x_2$, due to antisymmetrization of the wave function, while the bosonic state is maximal at this line, in this case due to symmetrization of the probability distribution. Via a $\sigma_z$ local rotation on the enlarged spinor, one can shift between these two behaviors.

We consider now $M=x_1^2x_2^2\sigma_1^x\sigma_2^x$ as our observable. The corresponding expectation values for bosonic and fermionic statistics and their correlations are given by
\begin{eqnarray}
&&\langle M \rangle_{\psi_b}=\Delta_1^2\Delta_2^2\cos^2(gt), \nonumber\\
&&\langle M \rangle_{\psi_f}=-\Delta_1^2\Delta_2^2\cos^2(gt), \nonumber\\
&&\langle M \rangle_{\psi_b, \psi_f}=0.
\end{eqnarray}
We plot in Fig.~\ref{JC}d the time-dependent dynamics of this observable for fermionic and bosonic statistics, which evolve with opposite phases, and thus make such an observable an appropriate candidate to differentiate between these two kind of behaviors. As expected, the correlations between fermions and bosons cancel due to the superselection rule.

Besides the example for $\psi_1=|\!\uparrow\rangle|0\rangle_1$, $\psi_2=|\!\downarrow\rangle|0\rangle_2$, we also investigate the case for higher initial Fock states, i.e. $n\neq m$. Here, we choose $n=3$ and $m=1$, without loss of generality. The corresponding expressions of bosonic and fermionic statistics at time $gt=\pi/2$ are reduced to
\begin{eqnarray}
\psi_b&=&\frac{i}{\sqrt{2}}\left[|3\rangle_1|0\rangle_2+|0\rangle_1|3\rangle_2\right]|\!\uparrow\rangle_1|\!\uparrow\rangle_2,\nonumber\\
\psi_f&=&\frac{i}{\sqrt{2}}\left[|3\rangle_1|0\rangle_2-|0\rangle_1|3\rangle_2\right]|\!\uparrow\rangle_1|\!\uparrow\rangle_2.
\end{eqnarray}

We can measure bosonic and fermionic statistics and their cross-correlation by taking $M=x_1^2x_2^2\sigma_1^x\sigma_2^x$ as an observable,
\begin{eqnarray}
&&\langle M \rangle_{\psi_b}=6\Delta_1^2\Delta_2^2\cos^2(gt)\cos^2(2gt),\nonumber\\
&&\langle M \rangle_{\psi_f}=-6\Delta_1^2\Delta_2^2\cos^2(gt)\cos^2(2gt), \nonumber\\
&&\langle M \rangle_{\psi_b, \psi_f}=0. 
\end{eqnarray}
Again, the cross-correlation vanishes because of the superselection rule.

\begin{figure}
\centering
\includegraphics[width=1\linewidth]{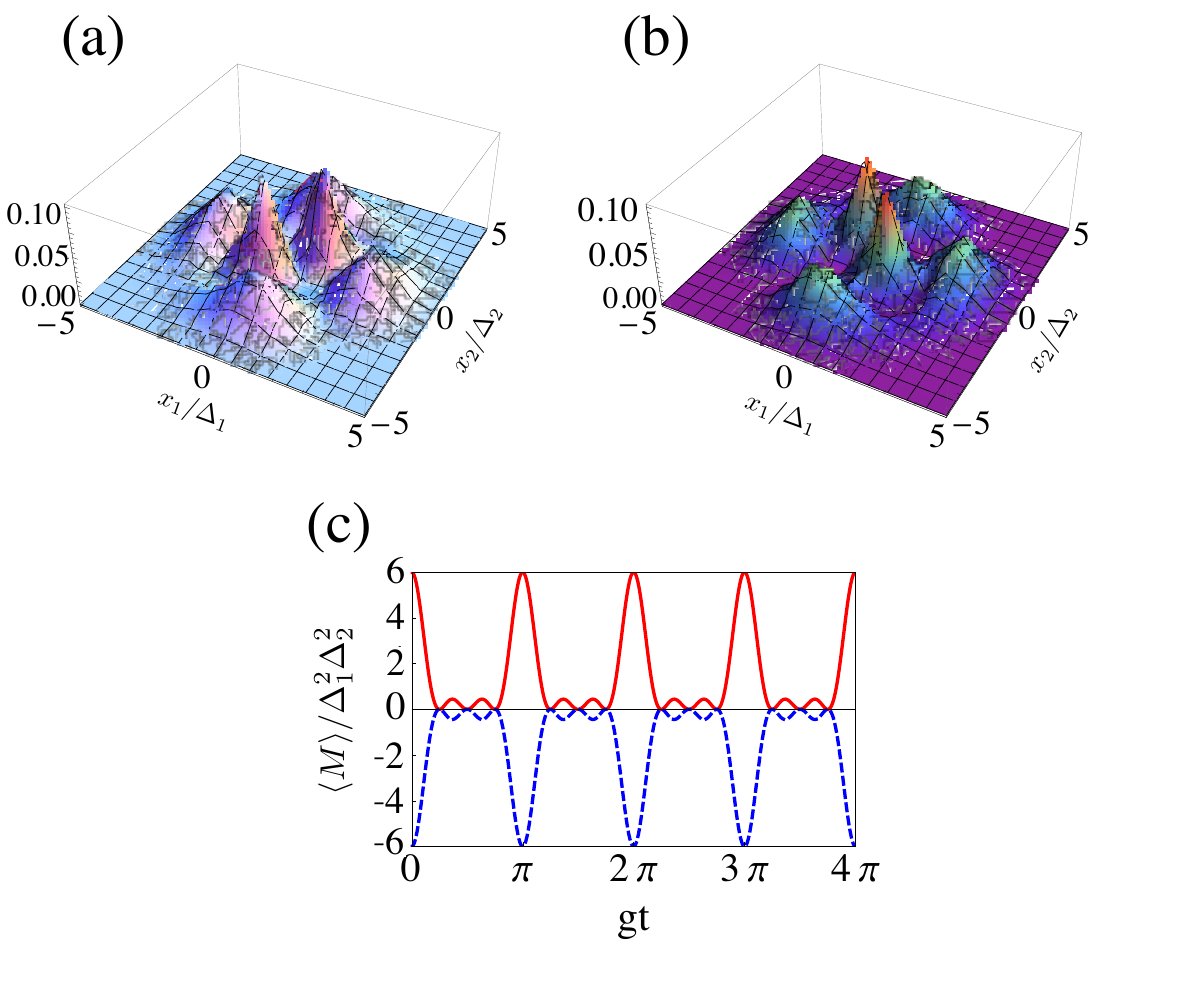}
\caption{(color online) Probability distributions $|\psi(x_1,x_2)|^2$ with higher initial Fock states for (a) Bose-Einstein and (b) Fermi-Dirac statistics, for $t=\pi/2g$, versus $x_1/\Delta_1$ and $x_2/\Delta_2$. (c) Time-dependent measurements for Bose-Einstein (solid red line) and Fermi-Dirac (dashed blue line) statistics versus $gt$ of the nontrivial correlation ${M=x_1^2x_2^2\sigma_1^x\sigma_2^x}$.}
\label{Jaynes Cummings Hamiltonian}
\end{figure}

\subsection{Motional-Spin Dynamics (quantum Rabi)}
\begin{figure}[]
\centering
\includegraphics[width=1\linewidth]{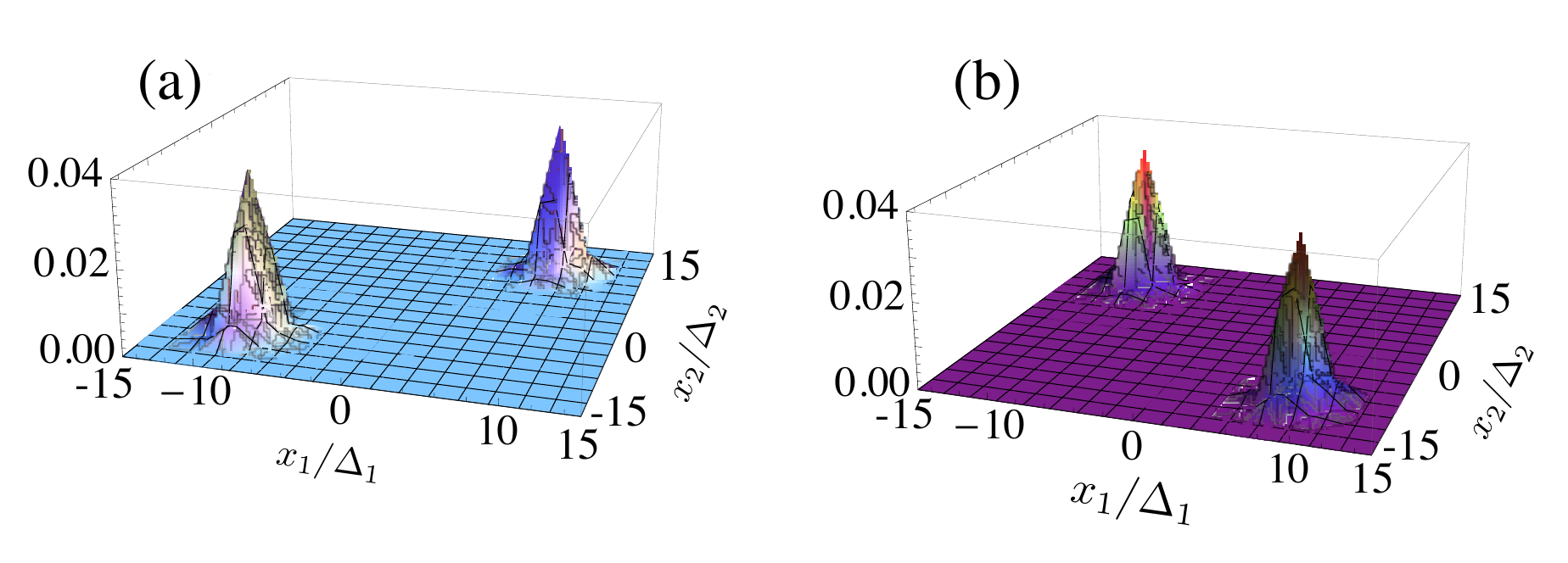}
\caption{(color online) Probability distributions $|\psi(x_1,x_2)|^2$ of all spins up subspace at time $t=1.4\pi/g$ for (a) Bose-Einstein and (b) Fermi-Dirac statistics, where $x_1$ and $x_2$ are in units of the position variances $\Delta_1$ and $\Delta_2$, respectively.}
\label{Rabi}
\end{figure}

We consider now a specific example of two particles containing spins and continuous degrees of freedom. The dynamics we choose for pedagogical purposes will be given by the quantum Rabi Hamiltonian in the simulated space, $H_{R}=g\sum\nolimits_{i=1}^{2}\sigma_i^x(a_i+a_i^{\dag})$, where $g$ is the coupling strength and $a_i$, $a_i^{\dag}$ are the annihilation and creation operators acting on the continuous degrees of freedom of the system. The corresponding Hamiltonian in the enlarged space is ${\tilde{H}_{R}=\mathbb{1}_2 \otimes H_{R}}$.  Proposals for simulating the quantum Rabi Hamiltonian in the ultrastrong coupling regime have already been made in trapped-ion setups~\cite{Rabi ion} and superconducting circuits~\cite{Rabi cQED}.
For the specific case of Gaussian wave functions as initial states of the continuous degree of freedom, we have ${\psi'(x_1,t=0)=|\!\uparrow\rangle_1|0\rangle_1}$, ${\psi''(x_2,t=0)=|\!\downarrow\rangle_2|0\rangle_2}$ with $\hat{x}_i=\Delta_i(a_i+a_i^{\dag})$, where we have made use of the fact that Fock state $| 0 \rangle$ is a Gaussian distribution centered in the origin of the phase space. The corresponding state for the enlarged spinor reads
\begin{eqnarray}
\Psi_0=\frac{1}{\sqrt{2}}\left(\begin{array}{cc}|\!\uparrow\rangle_1|0\rangle_1|\!\downarrow\rangle_2|0\rangle_2 \\ |\!\downarrow\rangle_1|0\rangle_1|\!\uparrow\rangle_2|0\rangle_2\end{array}\right),
\end{eqnarray}
which can be initialized via the generation of an entangled state in the spin degrees of freedom of
the two particles and the ancilla associated with the symmetrization spinor. Therefore, we can obtain the Bose-Einstein and Fermi-Dirac wave functions as a function of time with the corresponding mapping as previously explained. We plot in Fig.~\ref{Rabi} the probability distribution of both bosonic and fermionic cases in the subspace of all spins up, $|\!\uparrow\rangle_1|\!\uparrow\rangle_2$, for an evolution time $t=1.4\pi/g$. One can appreciate that the fermionic state cancels at the diagonal $x_1=x_2$, due to antisymmetrization of the wave function, while the bosonic state is maximal at this line, in this case due to symmetrization of the probability distribution. Via a $\sigma_z$ local rotation on the enlarged spinor, one can shift between these two behaviours.

\subsection{Three-Particle System Driven by Rabi Hamiltonian}
Now we consider the three-particle system, whose spinor in the embedding space is
\begin{equation}
\Psi=\frac{1}{\sqrt{6}}\left(\begin{array}{cccccccc}\psi'(q_1)\psi''(q_2)\psi'''(q_3) \\ \psi'(q_2)\psi''(q_1)\psi'''(q_3) \\ \psi'(q_3)\psi''(q_2)\psi'''(q_1) \\ \psi'(q_3)\psi''(q_1)\psi'''(q_2) \\ \psi'(q_1)\psi''(q_3)\psi'''(q_2) \\ \psi'(q_2)\psi''(q_3)\psi'''(q_1) \\ 0 \\ 0 \end{array}\right).
\end{equation}
We can generate both bosonic and fermionic states according to the protocol in the main article,
\begin{eqnarray}
\psi_b&=&\frac{1}{\sqrt{6}}[\psi'(q_1)\psi''(q_2)\psi'''(q_3)+\psi'(q_2)\psi''(q_1)\psi''(q_3)\nonumber\\
&&+\psi'(q_3)\psi''(q_2)\psi'''(q_1)+\psi'(q_3)\psi''(q_1)\psi'''(q_2)\nonumber\\
&&+\psi'(q_1)\psi''(q_3)\psi'''(q_2)+\psi'(q_2)\psi''(q_3)\psi'''(q_1)], \nonumber\\
\psi_f&=&\frac{1}{\sqrt{6}}[\psi'(q_1)\psi''(q_2)\psi'''(q_3)-\psi'(q_2)\psi''(q_1)\psi''(q_3)\nonumber\\
&&-\psi'(q_3)\psi''(q_2)\psi'''(q_1)+\psi'(q_3)\psi''(q_1)\psi'''(q_2)\nonumber\\
&&-\psi'(q_1)\psi''(q_3)\psi'''(q_2)+\psi'(q_2)\psi''(q_3)\psi'''(q_1)]. \nonumber\\
\end{eqnarray}

The measurements in the simulated space can be achieved via the transformation into the simulating space, by
\begin{eqnarray}
\langle M \rangle_{\psi_b}=&&\langle \Psi |\!\uparrow \rangle_c \langle \!\uparrow| \tilde M | \Psi\rangle=\langle \Psi_{sym}|(\mathbb{1}_2+\sigma_{1}^x)\nonumber\\
&&\otimes(\mathbb{1}_2+\sigma_{2}^x)\otimes(\mathbb{1}_2+\sigma_{3}^x)\otimes M|\Psi_{sym}\rangle/2,\nonumber\\
\langle M \rangle_{\psi_f}=&&\langle \Psi |\!\downarrow \rangle_c \langle \!\downarrow| \tilde M | \Psi\rangle=\langle \Psi_{asym}|(\mathbb{1}_2+\sigma_{1}^x)\nonumber\\
&&\otimes(\mathbb{1}_2+\sigma_{2}^x)\otimes(\mathbb{1}_2+\sigma_{3}^x)\otimes M|\Psi_{asym}\rangle/2,\nonumber\\
\langle M \rangle_{\psi_b,\psi_f}=&&\langle \Psi |\!\uparrow \rangle_c \langle \!\downarrow| \tilde M | \Psi\rangle=0.
\end{eqnarray}
\begin{figure}
\centering
\includegraphics[width=0.9\linewidth]{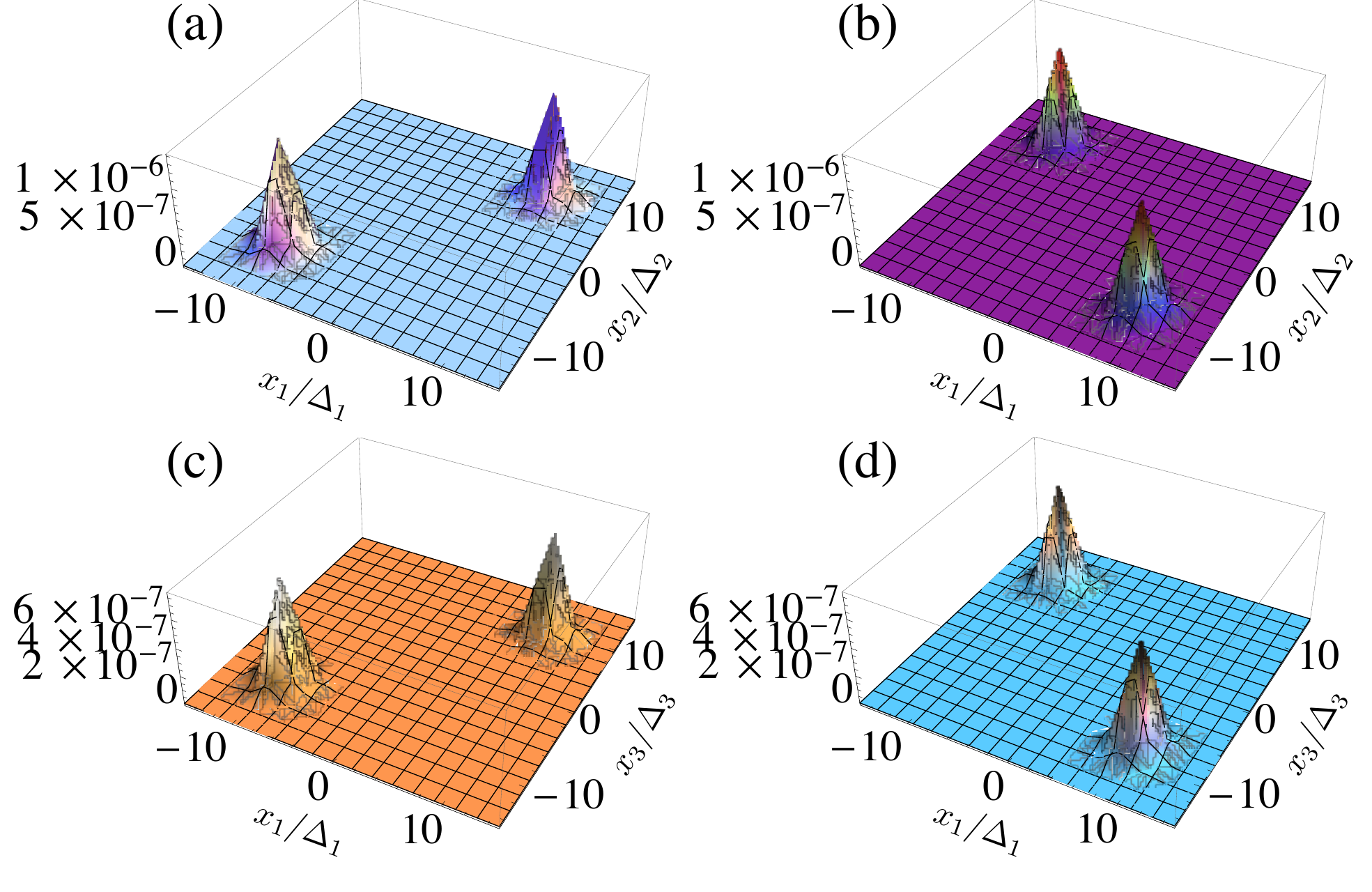}
\caption{(color online) Probability distributions $|\psi(x_1,x_2)|^2$ for (a) Bose-Einstein and (b) Fermi-Dirac statistics in the subspace of all spins up at $t=1.4\pi/g$ with $x_3=0$, versus $x_1/\Delta_1$ and $x_2/\Delta_2$. Probability distributions $|\psi(x_1,x_3)|^2$ for (c) Bose-Einstein and (d) Fermi-Dirac statistics in the subspace of all spins up at $t=1.4\pi/g$ with $x_2=0.5\Delta_2$, versus $x_1/\Delta_1$ and $x_3/\Delta_3$.}
\label{Rabi3}
\end{figure}
Here, we consider a three-particle system evolving under the dynamics given by the quantum Rabi model, $H_{R}=g\sum\nolimits_{i=1}^{3}\sigma_i^x(a_i+a_i^{\dag})$, where $g$ is the coupling strength and $a_i$, $a_i^{\dag}$ are the annihilation and creation operators acting on the continuous degrees of freedom of the system. Unlike the cases for two particles, two additional ancillary qubits are necessary for the embedding. The corresponding Hamiltonian in the enlarged space is ${\tilde{H}_{R}=\mathbb{1}_2\otimes\mathbb{1}_2\otimes\mathbb{1}_2 \otimes H_{R}}$. The initial states of the three particles in the simulated space are $\psi'(x_1,t=0)=|\!\uparrow\rangle_1|0\rangle_1$, $\psi''(x_2,t=0)=|\!\downarrow\rangle_2|0\rangle_2$,  and $\psi'''(x_3,t=0)=|\!\downarrow\rangle_3|1\rangle_3$, with $\hat{x_i}=\Delta_i(a_i+a^{\dag}_i)$. Its corresponding embedding spinor is expressed as
\begin{equation}
\Psi=\frac{1}{\sqrt{6}}\left(\begin{array}{cccccccc}|\!\uparrow\rangle_1|0\rangle_1|\!\downarrow\rangle_2|0\rangle_2|\!\downarrow\rangle_3|1\rangle_3 \\ |\!\uparrow\rangle_2|0\rangle_2|\!\downarrow\rangle_1|0\rangle_1|\!\downarrow\rangle_3|1\rangle_3 \\ |\!\uparrow\rangle_3|0\rangle_3|\!\downarrow\rangle_2|0\rangle_2|\!\downarrow\rangle_1|1\rangle_1 \\ |\!\uparrow\rangle_3|0\rangle_3|\!\downarrow\rangle_1|0\rangle_1|\!\downarrow\rangle_2|1\rangle_2 \\ |\!\uparrow\rangle_1|0\rangle_1|\!\downarrow\rangle_3|0\rangle_3|\!\downarrow\rangle_2|1\rangle_2 \\ |\!\uparrow\rangle_2|0\rangle_2|\!\downarrow\rangle_3|0\rangle_3|\!\downarrow\rangle_1|1\rangle_1 \\ 0 \\ 0 \end{array}\right).
\end{equation}
Thus, we can obtain the time-dependent wave functions of both Bose-Einstein and Fermi-Dirac statistics. Fig.~\ref{Rabi3}a-\ref{Rabi3}d show their probability distribution in the subspace of all spins up $|\!\uparrow\rangle_1|\!\uparrow\rangle_2|\!\uparrow\rangle_3$ at time $t=1.4\pi/g$, with $x_i$ fixed for one of the particles. It is clear that there is no probability of finding the particles in the diagonal $x_1=x_2$ (Fig.~\ref{Rabi3}b), and $x_1=x_3$ (Fig.~\ref{Rabi3}d) because of the wave function antisymmetrization for Fermi-Dirac statistics.

\subsection{Bose-Hubbard/Fermi-Hubbard switch}
Two prominent models in many-body systems are the Bose-Hubbard and the Fermi-Hubbard interactions. Here we give an application of our embedding for performing a switch between both models in the two-excitation subspace. A two-mode variant of the Bose-Hubbard Hamiltonian, in second quantization, can take the form $H=-t(b^\dag_1 b_2+b^\dag_2 b_1)+U b^\dag_1 b_1 b^\dag_2 b_2$, where in this simplified version both the hopping and interaction terms involve just the 1 and 2 modes. These obey standard commutation relationships, $[b_i,b^\dag_j]=\delta_{ij}$. On the other hand, the equivalent Fermi-Hubbard Hamiltonian takes the same form but with the corresponding modes obeying anticommutation relationships, $\{b_i,b^\dag_j\}=\delta_{ij}$. Being interested in the particle statistics, we may project onto the two-particle subspace, as the simplest non-trivial case to consider. The corresponding Hamiltonian for the Bose-Hubbard model, $H_b=-\sqrt{2} t(|2,0\rangle+|0,2\rangle)\langle 1,1 |_b+ |1,1\rangle_b(\langle 2,0| + \langle 0,2|) + U | 1, 1\rangle_b\langle 1, 1|_b$ differs from the one for the Fermi-Hubbard model, $H_f=U| 1, 1\rangle_f\langle 1, 1|_f$, given the distinct particle statistics. Moreover, in the bosonic case, $|1, 1\rangle_b \equiv (|1_1, 2_2\rangle+|1_2,2_1\rangle)/\sqrt{2}$ is the symmetrized superposition of having one particle in each mode, where $|1_i,2_j\rangle$ denotes particle 1 in mode $i$ and particle 2 in mode $j$, while the fermionic case $|1, 1\rangle_f \equiv (|1_1, 2_2\rangle-|1_2,2_1\rangle)/\sqrt{2}$ corresponds to the antisymmetrized superposition. Our embedding technique applied in this case considers the states $|s_1\rangle\equiv|2,0\rangle$, $|s_2\rangle\equiv|0,2\rangle$, $|s_3\rangle\equiv|1_1,2_2\rangle$, and $|s_4\rangle\equiv|1_2,2_1\rangle$. A superposition $|\psi\rangle=c_1|s_1\rangle+c_2|s_2\rangle+c_3|s_3\rangle+c_4|s_4\rangle$ written in spinor form, $\psi=[c_1, c_2, c_3, c_4]^T$, will then have the constraints in the Bose-Hubbard case, $c_3=c_4$, and in the Fermi-Hubbard case, $c_1=c_2=0$, $c_3=-c_4$. This allows one to obtain the bosonic(fermionic) wave function $\psi_b(\psi_f)$ via application of the corresponding transformation upon $\psi$, and the Hamiltonians $H_b, H_f$ can be straightforwardly embedded in this enlarged spinor similarly as before.

\section{Conclusions}

We have proposed the implementation of a quantum-particle-statistics switch with an embedding quantum simulator. We show that the simulated indistinguishable particles can change  {\it in situ}, from bosons to fermions and from fermions to bosons, during a quantum evolution, via their encoding in an enlarged spinor. Our proposal scales favorably in terms of ancillary-qubit resources with the number of particles and can be implemented with little additional effort in a conventional quantum simulator with mature platforms such as trapped ions, quantum photonics, or superconducting circuits.  Furthermore, additional exotic species as parastatistics can be encoded with small modifications of this protocol. The possibility to perform a switch of quantum particle statistics enhances the toolbox of quantum simulations, for unphysical operations as well as symmetry transformations, increasing their versatility.

\section*{Acknowledgements}
The authors acknowledge support from NSFC (11474193), the Shuguang Program (14SG35), the Program for Eastern Scholar, Specialized Research Fund for the Doctoral Program of Higher Education (2013310811003), the Basque Government with PhD grant PRE-2015-1-0394 and grant IT986-16, a UPV/EHU PhD grant, Ram\'{o}n y Cajal Grant RYC-2012-11391, UPV/EHU UFI 11/55, MINECO FIS2015-69983-P, the UPV/EHU project EHUA14/04 and the Chinese Scholarship Council (201506890077).

\end{document}